\documentclass[preprint,preprintnumbers,amsmath,amssymb]{revtex4}
\usepackage{graphicx}
\usepackage{dcolumn}
\usepackage{bm}
\usepackage{xcolor}

\begin{document}
	
\title{Coherent quantum work extraction of a relativistic battery as a probe for acceleration-induced Unruh thermality}
\author{Wei-Wei Zhang}
\author{Tian-Xi Ren}
\author{Yin-Zhong Wu}
\email{yzwu@mail.usts.edu.cn}
\author{Xiang Hao}
\email{xhao@mail.usts.edu.cn}
\affiliation{School of Physical Science and Technology, Suzhou University of Science and Technology, Suzhou, Jiangsu 215009, People's Republic of China}
\affiliation{Pacific Institute of Theoretical Physics, Department of Physics and Astronomy,
\\University of British Columbia, 6224 Agriculture Rd., Vancouver B.C., Canada V6T 1Z1.}
	
\begin{abstract}

We propose a physical scheme of a uniformly accelerated Unruh-DeWitt battery and utilize quantum work extraction as a probe to witness the thermal nature of the Unruh effect induced by the accelerated motion. By employing the open quantum system approach, we analyze the coherent and incoherent components of the ergotropy which is the maximum amount of quantum work extraction of the relativistic battery driven by a coherent field. It has been proved that the steady values of coherent quantum work extraction in the asymptotic condition is only determined by the acceleration-dependent Unruh temperature. The asymptotic behavior of coherent ergotropy can demonstrate the thermal nature of the Unruh effect with respect to the Kubo-Martin-Schwinger condition. Under the circumstance of the Unruh effect, we explore the effect of the phase of the coherent charging field on the dynamics of coherent ergotropy when the battery approaches to the same thermal equilibrium state. The variation in the phase of the coherent driving field can improve the energy storage capacity of a relativistic battery. From viewpoint of energy transfer, the relativistic battery is helpful to examine the Unruh thermality.
		
\end{abstract}
	
\maketitle
	
\section{Introduction}
	
Arising from black hole evaporation, the Unruh effect provided the evidence that a uniformly accelerated observer in Minkowski vacuum may detect a Planckian spectrum of particles with a thermal temperature proportional to the acceleration \cite{Unruh1976}. The conclusion can be extended to many phenomena including Hawking radiation of black holes \cite{Hawking1975} or particle excitation in curved spacetime \cite{Gibbons1977,XLiu2016}. The Unruh effect has been explored by an accelerated Unruh-DeWitt(UDW) detector moving along a constant acceleration trajectory \cite{Crispino2008}. However, observing the thermal nature of the Unruh effect remains challenging in present experimental setups \cite{Bell1983,Cardi2025}. According to the thermalization theorem, an accelerated detector experiences the loss of information ensured by the Kubo-Martin-Schwinger(KMS) condition which does not directly imply a Planck distribution \cite{Takagi1986}. For instance, in a massive field of a high dimensional spacetime, the response spectrum of the detector may differ distinctly from the Planckian shape \cite{Sriramkumar2003,Arrechea2021}. Due to the open question of Unruh thermality, some recent studies focus on the combination of relativity and quantum information theory \cite{Mann2012}. Some facets of quantum nonlocality\cite{Martin2015,Han2018,Wang2025}, geometric phase\cite{Yu2012}, and quantum parameter estimation\cite{Hao2019,Patterson2023} have been used to evaluate the thermal nature of the Unruh effect. These methods allow for a comprehensive examination of the connection between quantum correlation and Unruh effect. Our work aims to determine the thermal nature of the Unruh effect by examining quantum energy transfer of the detector.

In this paper, we put forward a relativistic UDW battery and explore the coherent and incoherent parts of quantum work extraction as a witness to Unruh thermality. The physical essence of a quantum battery reveals the fact that quantum resources can be used to transfer and store energy from the coupling to an external charger \cite{Alicki2013,Millen2016,Binder2015,Hao2023,Rossini2019,Dou2023,Francica2020,Niu2024}. If a quantum battery is in a non-passive state, quantum work can be extracted via a unitary cyclic operation. The maximal amount of quantum work extraction is defined as ergotropy for evaluating the storage performance. The ergoropy can be divided into an incoherent and coherent component which correspond to a coherence-preserving operation and coherence-consuming one respectively \cite{Francica2020}. The coherent quantum work extraction is dependent on nonclassical features of a quantum battery. In our scheme, the relativistic UDW battery is viewed as a uniformly accelerated detector which is driven by a coherent field and coupled to scalar vacuum fields in Minkowski spacetime. We treat the battery as the open quantum system immersed in the environment of fluctuating vacuum scalar fields. The dynamics of the relativistic battery is modulated by the Unruh decoherence. Considering the role of quantum coherence, we attempt to probe Unruh thermal nature by exploiting the dynamics of the coherent quantum work extraction. In the asymptotic condition, the coherent ergotropy for steady states can demonstrate the thermal nature of the Unruh effect.

The reasons for investigating Unruh thermality by means of coherent ergotropy lie in the following factors. Firstly, as an operational quantification of energy transfer between the detector and Minkowski vacuum, coherent contribution to quantum work extraction may incorporate the special feature of the response spectrum for the UDW battery. In some sense, the coherent ergotropy facilitates the observation of the quantum traits of the Unruh effects. Moreover, the dynamics of the quantum work extraction is governed by quantum coherence of the moving relativistic battery. Many researches have demonstrated that the thermal nature of the Unruh effect captures quantum correlations across Rindler horizon \cite{Ahmadi2014,Zhao2020,Liu2021,Benatti2004}. The nonclassical feature of Unruh thermality is connected with vacuum quantum fluctuations in relativistic thermodynamics. The behavior of coherent ergotropy allows us to probe quantum properties of Minkowski vacuum fields, thus providing the reliable evidence for the thermal nature of the Unruh effect.

The paper is organized as follows. In Sec. II, we suggest a physical model of a relativistic quantum battery and introduce the coherent and incoherent contributions to quantum work extraction, from viewpoint of relativistic quantum thermodynamics. We discover the relationship between the response spectrum and quantum work extraction. In Sec. III, we study the evolutions of coherent ergotropy as function of the acceleration-induced Unruh temperature and a relative phase of a coherent charging field. The asymptotic behavior of the coherent component of quantum work extraction manifest the Unruh thermality. Changing the phase of the coherent field helps for the optimization of coherent ergotropy and improving the performance of the relativistic battery. Finally, in Sec. IV, we give our conclusions and discussions.

\section{The relativistic UDW battery in Minkowski vacuum}

We propose a relativistic quantum battery model in Minkowski spacetime. The physical model of the battery is provided by a uniformly accelerated two-level atom which is simultaneously driven by a coherent field. The battery can be treated as an open quantum system interacting with a bath of fluctuating quantum scalar fields in Minkowski spacetime. It can be referred to as a relativistic UDW battery, which is charged by a coherent field and influenced by Minkowski vacuum fields. The Hamiltonian of a static battery is expressed as ${H_0} = {\omega_0}{\sigma ^ + }{\sigma ^ - } $, where ${\omega_0} $ represents the energy gap between the excited state $ \left| e \right\rangle $ and ground state $\left| g \right\rangle $, and ${\sigma ^ \pm }$ are the raising and lowering operators, respectively. At the beginning of a charging process, the battery in its ground state $\left| g \right\rangle $ denotes a depleted status. During the charging interval $\tau $, an external coherent field is used to charge the battery via the dipole interaction between the atom and field in the resonant condition.

The external field with the frequency $\omega$ is described as $\mathbf{E}(t) = {\mathbf{E}_0}\cos (\omega t + \varphi ) $, where ${\mathbf{E}_0} $ and $\varphi$ represent the amplitude and initial phase of the field, respectively. The Hamiltonian of the driven battery can be written as $\tilde{H} = H_0 - \mathbf{d}\cdot \mathbf{E}$ where the second term denotes the dipole interaction between the external field and battery. The parameter $\mathbf{d}=\mathbf{d}_0(\sigma^+ +\sigma^-)$ represents the electric dipole moment of a two-level atom. In the interaction picture, the transformed Hamiltonian $H_b=e^{iH_0 t} \tilde{H} e^{-iH_0 t}$ can be given by $H_b=\frac {\Omega}2[\sigma ^+(e^{i(\omega_0+\omega) t+i\varphi} +e^{i(\omega_0-\omega) t-i\varphi} )+ \sigma ^- (e^{-i(\omega_0+\omega) t-i\varphi}+e^{-i(\omega_0-\omega) t+i\varphi})]$ where we can neglect the high-frequency oscillating parts $e^{\pm i(\omega_0+\omega)t}$ in the rotating wave approximation. Here $\Omega=-\mathbf{d}_0\cdot \mathbf{E}_0$ represents the effective coupling strength. Considering the resonance condition of $\omega=\omega_0$, we can obtain the definite expression of ${H_b} = \mu(t)\frac {\Omega}{2}({\sigma ^+}{e^{- i\varphi}} + {\sigma ^-}{e^{i\varphi}} )=\mu(t)\frac {\Omega}{2}(\mathbf{n}\cdot \bm{\sigma})$ where $\mathbf{n} = (\cos \varphi ,\sin \varphi ,0)$ is the unit vector. The function $\mu(t)=1(0\leq t \leq \tau)$ is applied to control the charging process. One simple case of $\varphi=0$ has been studied in the previous work \cite{Mukherjee2024,Gemme2022}.

In an inertial condition, the battery obeys the evolution of $\rho (\tau) = U(\tau ){\rho _0}{U^\dag }(\tau )$ where $U(\tau ) ={\mathcal T}\exp [ - i\int_0^\tau  {ds{H_b}(s)} ]$ is the unitary cyclic operation and $ \mathcal T$ representing the time-ordering operator and ${\rho _0}$ is an initial state. Any unitary operation of the battery is determined by the atom-field interaction. According to \cite{Francica2020}, the external field phase will have an effect on quantum work-extraction operations $U(\tau)$, which can vary the incoherent and coherent components of the ergotropy. This fact motivates us to study the role that the phase may play. According to \cite{Alicki2013}, the maximal amount of quantum work extraction, i.e., the ergotropy, can be optimized over all unitary operators $\{ U(\tau ) \}$ as
\begin{equation}
	\label{eq:(1)}
    \mathcal{W}(\tau ) = Tr[{H_0}\rho (\tau )] - \mathop {\min }\limits_{\{ U\} } Tr[U\rho (\tau ){U^\dag }{H_0}],
\end{equation}
where $Tr[{H_0}\rho (\tau )]$ is the total energy of the battery and the second term represents the minimum energy achievable under any unitary transformation. Due to the optimal transformation ${U_\sigma }$, the battery evolves into a passive state $\rho_{\sigma}=\sum_{j}\varrho_j|\epsilon_j\rangle \langle \epsilon_j| $. According to the theorem of \cite{Alicki2013,Pusz1977,Lenard1978}, a state is passive when it commutes with the Hamiltonian of the battery $H_0=\sum_j \epsilon_j |\epsilon_j\rangle \langle \epsilon_j|$ and its eigenvalues are non-increasing with the energy. Given $\rho(\tau)$ there is a unique passive state $\rho_{\sigma}$ which minimize $Tr[U\rho(\tau) U^{\dag}H_0]$ by the optimized operation. In this case, no work can be extracted from the passive state. Here, $|\epsilon_j\rangle $ is the energy level state of $H_0$ with the corresponding energy $\epsilon_j$ in the increasing order, $\epsilon_j<\epsilon_{j+1}$ and $\varrho_j$ is the eigenvalues of $\rho(\tau)=\sum_j \varrho_j |\varrho_j\rangle \langle \varrho_j|$ in the decreasing order.

To demonstrate the connection between the ergotropy $\mathcal{W}$ and the coherence $\mathcal{C}$, the ergotropy can be sequently divided into two contributions, where the incoherent component $\mathcal{W}_i$ is firstly accomplished by a coherence-preserving unitary operation, and the residual coherent ergotropy $\mathcal {W}_r$ is extracted via a coherence-consuming unitary operation. According to \cite{Francica2020}, the incoherent ergotropy $\mathcal{W}_i$ is the maximal amount of work extraction without changing quantum coherence of the battery. In this case, over all incoherent cyclic unitary operations $\{ V \}$, the definition of $\mathcal{W}_i$ is written as $\mathcal{W}_i(\rho) =\max_{V} \{ \mathrm{Tr}[{H_0}\rho] - \mathrm{Tr}[V\rho{V^\dag }{H_0}] \}$. The coherence preservation condition $\mathcal{C}(\rho ) = \mathcal{C}(V\rho {V^\dag })$ holds for any quantum state. The optimal coherence-preservation transformation is provided by ${V_\pi }$, where $\pi$ denotes the permutation of energy basis up to irrelevant phase factors. In the eigenvalues of $H_0$ are arranged in an descending order, ${V_\pi }$ is applied to rearrange the population of $\rho$ in an ascending order. The special state after the optimal incoherent extraction is expressed as $\rho_{\pi}=V_\pi \rho V_\pi^{\dag}$ and maintains the coherence of $\rho$ but less energy. For example, the optimal incoherent operator for a two-level battery is obtained by $V_{\pi}=|e\rangle \langle g|-|g\rangle \langle e|=\sigma^{+}-\sigma^{-}$ which swap the order of population in energy basis \cite{Niu2024}. The second step is to apply a coherence-consuming operation which converts the special state $\rho_{\pi}$ into passive state $\rho_{\sigma}$. During this period, the residual coherent ergotropy $\mathcal{W}_r$ is extracted as
\begin{equation}
\label{eq:(2)}
\mathcal{W}_r=\mathcal{W}-\mathcal{W}_i.
\end{equation}
The coherent ergotropy can be understood as that part of extractable work arising from the presence of coherence in the initial state. From the perspective of quantum thermodynamics, this tool will provide us a new way to measuring quantum traits of the Unruh effect.

In Minkowski spacetime, the interaction between the accelerated battery and fluctuating vacuum fields leads to Unruh decoherence. During the charging process, the battery moves along a trajectory $x(\tau ) = (t, \vec{x})$ with the uniform acceleration of $a$. The combined Hamiltonian for the battery and scalar field is expressed as
\begin{equation}
	\label{eq:(3)}
	H = {H_b} + {H_f} + {H_I},
\end{equation}
where ${H_f}$ is the Hamiltonian of the scalar field $\Phi (x(\tau ))$ satisfying the standard Klein-Gordon equation in Minkowski spacetime. The part ${H_I} = \mu(\tau) \lambda ({\sigma ^ + } + {\sigma ^ - })\Phi (x(\tau ))$ describes the interaction between UDW battery and vacuum scalar fields where $\lambda $ is the coupling constant. The weak coupling condition of $\lambda\ll \Omega$ is considered. In fact, the energy storage of the relativistic UDW battery relies on both external coherent field and fluctuating vacuum fields in Minkowski spacetime. In the weak coupling condition, the dynamics of the UDW battery can be obtained by an open quantum system approach. At $\tau=0$, the combined system is in a separable state, ${\rho _{\mathrm{tot}}}(0) = |g \rangle\langle g|  \otimes | 0 \rangle \langle 0|$, where $\left| 0 \right\rangle $ represents the vacuum state of the scalar field in Minkowski spacetime. In the frame of the moving battery, the reduced density matrix $\rho (\tau )$ of the battery can be obtained through the quantum master equation \cite{Benatti2004}, which is written in the Kossakowski-Lindblad form of
\begin{align}
\label{eq:(4)}
\frac {\partial}{\partial \tau} \rho(\tau)&\;=\;-i[H_{b}^{(\mathrm{eff})}, \rho(\tau)]+\frac 12\sum_{i,j=1}^{3}a_{ij}\mathcal{D}_{ij}[\rho(\tau)], \\
a_{ij}&\;=\;A\delta_{ij}-iB\varepsilon_{ijk}n_k+Cn_in_j, \nonumber \\
A&\;=\; \frac {\lambda^2}{2}[\mathcal{G}(\Omega)+\mathcal{G}(-\Omega)],\;B= \frac {\lambda^2}{2}[\mathcal{G}(\Omega)-\mathcal{G}(-\Omega)],\;C=\lambda^2 \mathcal{G}(0)-A.\nonumber
\end{align}
where the dissipator $\mathcal{D}_{ij}(\rho)=2\sigma_j\rho\sigma_i-\sigma_i\sigma_j\rho-\rho\sigma_i\sigma_j$ arises from the dissipation and decoherence induced by the environment. $\{ \sigma_j,(j=1,2,3)\}$ are the three components of Pauli operators. The Kossakowski matrix $a_{ij}$ can be explicitly resolved. The detailed and independent derivation of Eq. (4) is provided in Appendix A. By introducing the Wightman function of scalar field $ G^{+}(x-x')=\langle 0|\Phi \big( x(\tau) \big)\Phi \big( x'(\tau^{'}) \big)|0\rangle=\frac {1}{4\pi^2[|\vec{x}-\vec{x}'|^2-(t-t'-i\epsilon)]}$, we can derive its Fourier transform
\begin{equation}
\label{eq:(5)}
\mathcal{G}(\Omega)=\int_{-\infty}^{\infty} \mathrm{d}\Delta \tau \cdot e^{i\Omega\Delta \tau} G^{+}(\Delta \tau).
\end{equation}
The Hilbert transform of the Wightman function is given by $\mathcal{K}(\Omega)=\frac {\mathcal{P}}{\pi i}\int_{-\infty}^{\infty} \mathrm{d}\omega\frac {\mathcal{G}(\omega)}{\omega-\Omega}$
where $\Delta \tau=\tau-\tau'$ and $\mathcal{P}$ denotes the principle value. The effective Hamiltonian is given by $H_{b}^{(\mathrm{eff})}=\frac {1}{2}\Omega^{'}(\sigma^{+}+\sigma^{-})$ with $\Omega^{'}=\Omega+i\lambda^2[\mathcal{K}(-\Omega)-\mathcal{K}(\Omega)]$ representing the effective coupling. The interaction with external scalar field would have an effect on the Lamb shift. In the case of weak couplings, $\lambda^2 \ll \Omega$, we can neglect the Lamb shift in the following.

For the UDW battery with a constant acceleration $a$ in Minkowski spacetime, we consider the motion trajectory in the form of,
\begin{equation}
\label{eq:(6)}
x(\tau)=\left(\frac 1a \sinh(a\tau),\frac 1a \cosh(a\tau),0,0  \right).
\end{equation}
With respect to the UDW battery along the trajectory, the field Wightman function ${G^ + }(\Delta \tau ) =-\frac{{{a^2}}}{{16{\pi ^2}}}\frac{1}{{{{\sinh }^2}(\frac{{a\Delta \tau }}{2}-i\epsilon)}}$ obeys the KMS condition, i.e., $G^+(\Delta\tau) = G^+(\Delta\tau+ i\beta)$. Equivalently, in the frequency space, the KMS condition can be demonstrated by $\mathcal{G}(\Omega)=\frac {\Omega}{2\pi(1-e^{-\beta \Omega})}= e^{\beta \Omega}\mathcal{G}(-\Omega)$ where $\beta=\frac {1}{T_U}=\frac {2\pi}{a}$ represents the Unruh temperature.
To mathematically describe the state of the battery, we typically use the Bloch vector $ \mathbf{r} (\tau )$ to represent the evolved density matrix $\rho (\tau ) = \frac{{{\rm I} + \sum\nolimits_j {{r_j}(\tau ){\sigma _j}} }}{2}$ where $r_j=\mathrm{Tr}(\sigma_j \rho)$ is the $j$th-component of the Bloch vector.

According to the Lindblad master equation, the dynamics of the battery can be transformed into the Bloch equation,
\begin{equation}
	\label{eq:(7)}
    \frac{d}{{d\tau }}\mathbf{r}(\tau ) =  - 2\bm{\mathcal{H}} \cdot \mathbf{r}(\tau ) + \bm{\chi}(\tau),
\end{equation}
where $\bm{\mathcal{H}}$ is a decaying matrix and given by
\begin{equation}
	\label{eq:(8)}
	\bm{\mathcal{H}} = \left( {\begin{array}{*{20}{c}}
		{2A + C{{\sin }^2}\varphi }&{ - C\cos \varphi \sin \varphi }&{ - \frac{\Omega }{2}\sin \varphi }\\
		{ - C\cos \varphi \sin \varphi }&{2A + C{{\cos }^2}\varphi }&{\frac{\Omega }{2}\cos \varphi }\\
		{\frac{\Omega }{2}\sin \varphi }&{ - \frac{\Omega }{2}\cos \varphi }&{2A + C}
\end{array}} \right).
\end{equation}
Here the inhomogeneous vector $\bm{\chi}(\tau )= - 4B{(\cos \varphi ,\sin \varphi ,0)^{\rm T}}$. To facilitate the calculation, we use the quantum channel to describe the dynamics of the battery in the form of,
\begin{equation}
	\label{eq:(9)}
    \mathbf{r}(\tau ) = \bm{\Gamma}(\tau) \cdot\mathbf{r}(0) + \mathbf{\Lambda}(\tau ),
\end{equation}
where the mapped matrix of the quantum channel $\bm{\Gamma} (\tau )=\exp (-2\bm{\mathcal{H}}\tau)$ is expressed as
\begin{equation}
	\label{eq:(10)}
   \bm{\Gamma} (\tau ) = {e^{ - 4A\tau }} \begin{pmatrix}
		\cos^2 \varphi  + \sin^2\varphi f_c&(1 - f_c)\sin \varphi \cos \varphi &\sin \varphi f_s\\
		(1 - f_c)\sin \varphi \cos \varphi &\sin^2\varphi  + \cos^2\varphi f_c & -\cos \varphi f_s\\
		-\sin \varphi f_s &\cos \varphi f_s &f_c
        \end{pmatrix}.
\end{equation}
and $\bm{\Lambda} (\tau ) = \frac{1}{2}[I - \bm{\Gamma} (\tau )]{\bm{\mathcal{H}}^{ - 1}}\cdot \bm{\chi}$ is the mapped vector. Here $f_c=e^{-2C\tau}\cos (\Omega \tau )$ and $f_s=e^{-2C\tau}\sin (\Omega \tau )$ are the parameters as function of the battery-field coupling. The Unruh channel can be established by means of the mapped matrix and vector.

\section{The behavior of coherent and incoherent ergotropy in Unruh reservoir}

This work aims to investigate the coherent contribution to quantum work extraction of the relativistic UDW battery, which is used to quantitatively examine Unruh thermality from viewpoint of energy transfer. In regard to the steady state only relevant to the Unruh temperature, we explore the asymptotic behavior of coherent and incoherent ergotropy to reveal the thermal nature of the Unruh effect. Furthermore, we will study the effect of the phase of the coherent driving field on coherent ergotropy. In this section, we assume that the UDW battery initially evolves from the ground state.

Assuming the UDW battery is prepared in the ground state $\left| g \right\rangle $, we can obtain the evolved state in the Bloch vector,
\begin{equation}
	\label{eq:(11)}
	\mathrm{r}(\tau ) = \left( \begin{array}{l}
		\gamma[\cos \varphi ({e^{ - 4A\tau }} - 1)] - {e^{ - 2(2A + C)\tau }}\sin \varphi \sin (\Omega \tau )\\
		\gamma[\sin \varphi ({e^{ - 4A\tau }} - 1)] + {e^{ - 2(2A + C)\tau }}\cos \varphi \sin (\Omega \tau )\\
		\frac {\gamma}{2}[{e^{ - 4A\tau }}\sin (2\varphi )\sin (\Omega \tau )(1 - {e^{ - 2C\tau }})] - {e^{ - 2(2A + C)\tau }}\cos (\Omega \tau )
	\end{array} \right).
\end{equation}
where the ratio $\gamma=\frac {B}{A}=\tanh(\frac {\pi \Omega}{a})$ is only dependent on the Unruh temperature owing to the KMS condition. In the asymptotic limit of $\tau \rightarrow \infty$, the steady state of the battery is written as $\mathbf{r}_s=-\gamma(1,0,0)^{\mathrm{T}}$ associated with the Unruh temperature. It is worth noticing that the time-dependent evolution of the battery is also related to the phase of the coherent driving field. The following context will focus on the physical connection between Unruh thermality and coherent ergotropy.

By employing the correlation function ${G^ + }(\Delta \tau )$ of the vacuum field along this trajectory, we firstly obtain the decaying parameters of the Unruh channel,
\begin{align}
\label{eq:(12)}
A=&\; \frac {\lambda^2\Omega}{4\pi}\coth (\frac {\pi \Omega}{a}),\;B=\frac {\lambda^2\Omega}{4\pi},\;C=\lambda^2 \mathcal{G}(0)-A.
\end{align}

In our work, to obtain the maximum work extracted from the quantum state, it is necessary to transform the quantum state $\rho (\tau )$ into a passive state $\rho_{\sigma}=U_{\sigma}\rho(\tau)U_{\sigma}^{\dag}=\sum_{j}\varrho_j|\epsilon_j\rangle \langle \epsilon_j|$ through an optimal unitary operation $U_{\sigma} $ \cite{Alicki2013}.

Here, $\varrho_{1,2}=\frac {1\pm |\mathbf{r}(\tau)|}2$ is the eigenvalue of $\rho (\tau )$ which satisfies the condition $\varrho_2=\frac {1-r}{2} \le \varrho_1=\frac {1+r}{2}$ and $|\epsilon_1\rangle=|g\rangle,|\epsilon_2\rangle=|e\rangle $ is the corresponding eigenvector in the increasing order. Here, $r=|\mathbf{r}(\tau)|$ represents the magnitude of the Bloch vector. This optimal unitary operation is defined as $U_{\sigma}=\sum_{j=1,2}|\epsilon_j\rangle \langle \varrho_j|$, where $|\varrho_j\rangle$ is the eigenvector corresponding to the state $ \rho(\tau)$. In the Hilbert space spanned by $\{|\epsilon_1\rangle,|\epsilon_2\rangle \}$, we can obtain $|\varrho_{1}\rangle=\sqrt{\frac {r+r_3}{2r}}|\epsilon_1\rangle+\frac {r_1+ir_2}{\sqrt{2r(r+r_3)}}|\epsilon_2\rangle$ and $|\varrho_{2}\rangle=\frac {r_1-ir_2}{\sqrt{2r(r+r_3)}}|\epsilon_1\rangle-\sqrt{\frac {r+r_3}{2r}}|\epsilon_2\rangle$, respectively.  Therefore, the matrix form of the optimal unitary transformation is,
\begin{equation}
\label{eq:(13)}
U_{\sigma}=\sqrt{\dfrac {r+r_3}{2r}}(|\epsilon_1\rangle \langle \epsilon_1|-|\epsilon_2\rangle \langle \epsilon_2|)+\dfrac {1}{\sqrt{2r(r+r_3)}}[(r_1-ir_2)|\epsilon_1\rangle \langle \epsilon_2| + H.c.],
\end{equation}
where $H.c.$ denotes the Hermitian conjugate term of the former term. By the definition of maximum extractable work, the total energy of the battery can be calculated as $E(\tau)=\mathrm{Tr}[H_0\rho(\tau)]=\frac {\omega_0(1+r_3)}2$, and the optimal energy obtained after all unitary transformations is $\mathrm{Tr}[U_{\sigma}\rho(\tau)U_{\sigma}^{\dag}H_0]=\mathrm{Tr}[\rho_{\sigma}H_0]=\sum_{j=1,2}\varrho_j \epsilon_j= \frac {\omega_0(1-r)}2$. Here, $ \epsilon_j $ are the two eigenvalues of $ {H_0}$. At this point, the maximum extractable work can be calculated as $\mathcal{W}=\frac {\omega_0(r+r_3)}2$. For simplification, we define a scaled ergotropy  $\xi(\tau)=\frac {\mathcal{W}}{\omega_0}$,
\begin{equation}
	\label{eq:(14)}
	\xi(\tau)=\frac {1}{2}[r(\tau)+r_3(\tau)].
\end{equation}

In order to verify the coherent ergotropy, we firstly obtain the incoherent component of the maximal quantum work extraction via incoherent operations. For a two-level battery, we exploit the optimized coherence-preserving unitary operation $V_{\pi}=\sigma^{+}-\sigma^{-}$ to obtain the incoherent ergotropy ${\xi _i}=\frac {\mathcal{W}_i}{\omega_0}$ in the form of
\begin{equation}	
\label{eq:(15)}
	\xi _i(\tau)= \max \{r_3(\tau), 0 \}.
\end{equation}
Then, we can write the scaled coherent ergotropy as $\xi_r(\tau)=\frac {\mathcal{W}_r}{\omega_0}=\xi-\xi_i$ which is just the work extraction from quantum coherence of the UDW battery. In the condition of no external coherent field, i.e., $\Omega=0$, the asymptotic value of the incoherent ergotropy $\xi_i^{(s)}=0$ will vanish due to the result of $r_3(\infty)=-r(\infty)<0$ and the total ergotropy is also null. This is the reason that the battery without driving will arrive at the Unruh thermal equilibrium state which is a passive state. When the external coherent field is applied, the steady state with a certain value of quantum coherence is a non-passive one and the corresponding ergotropy is just determined by the coherent component of quantum work extraction. The ergotropy distribution in the asymptotic limit is given by,
\begin{equation}
\label{eq:(16)}
\xi^{(s)}=\xi_r^{(s)}=\frac {\gamma}2.
\end{equation}
In the following calculation, we will work with dimensionless parameters by scaling the acceleration and proper time as $\tilde{\tau}=\lambda^2 \Omega \tau$ and $\tilde{a}=\frac {a}{\Omega}$. For convenience, we continue to term $\tilde{a}$, $\tilde{\tau}$ as $a$, $\tau$, respectively. Fig. 1 illustrates that the asymptotic coherent ergotropy is just determined by the acceleration-induced Unruh temperatures. The coherent ergotropy monotonically decreases with increasing the acceleration. It is known that the Unruh effect can deteriorate quantum coherence of the accelerated detector. Unruh decoherence gives rise to the asymptotic behavior of the coherent ergotropy. This evidence demonstrates that the coherent ergotropy reflects the thermal nature of the acceleration-induced Unruh effect.

The detailed evolution of the coherent ergotropy for a certain acceleration is also studied when the phase of the external coherent field is varied. In Fig. 2(a), we demonstrate that coherent ergotropy can be oscillatorily promoted to one stable value. During the early evolution, the coherent ergotropy can be obviously modulated by the external field phase. The phase variation contributes to the improvement of the coherent ergotropy. This is the reason that Unruh decoherence has weak effects on quantum coherence in the early stage while quantum coherence can be adjusted by the interaction between the battery and coherent field. When the battery gradually arrives at the steady state, the Unruh effect plays a dominate role in the dynamics of the UDW battery.

To further investigate the effect of the external field phase on the coherent ergotropy, we show that the incoherent ergotropy is determined by $\xi_i(\tau)=\max \{\frac {\gamma}2 e^{-4A\tau}[\sin(2\varphi)(\sin \Omega \tau -f_s)-f_c], 0\}$ which is is dependent on the sine function of the phase. This point provides the reason that the oscillation behavior of the ergotropy occurs with the phase. From Fig. 2(b), it is seen that the maximal value of the coherent ergotropy is obtained by the optimal selection of the field phase while the minimal value of incoherent ergotropy arrives with respect to the same phase.

On the other hand, we explore the behavior of the coherent ergotropy due to the joint counterbalance of coherence driving field and Unruh thermality. In Fig. 3, it is clearly seen that the impacts of the coherent field phase on the coherent ergotropy are significant at low Unruh temperatures with small accelerations. It is reasonable that quantum deoherence at a low Unruh temperature has a weak influence on quantum work extraction. However, with increasing the acceleration, we find that coherent ergotropy is nearly independent of the external field phase. Under the circumstance of high Unruh temperatures with large accelerations, Unruh thermality primarily suppress the dynamics of the coherent ergotropy. In a word, the interaction between the battery and the fluctuating vacuum field has much more effects on the quantum work extraction in the condition of large accelerations.

\section{Discussion}
We have proposed the physical model of a relativistic UDW battery charged by an external coherent field and fluctuating vacuum fields in Minkowski spacetime. We have demonstrated that the coherent contribution to the maximal work extraction can be viewed as a witness to the thermal nature of the acceleration-induced Unruh effect. We have treated the UDW battery as the open quantum system. By means of Unruh channel, we effectively describe the dynamics of the UDW battery driven by the external coherent field with the controllable phase. The time-dependent behavior of the coherent ergotropy is closely related to the acceleration-induced Unruh temperature. We find that the asymptotic behavior of the coherent ergotropy is monotonically decreased with the acceleration. Such phenomena only dependent on acceleration can explain the the thermal nature of the Unruh effect. During the early evolution, the coherent ergotropy can be enhanced with the variation of the external field phase. In the condition of low accelerations, the modulation effect of the phase on the quantum work extraction is manifest. The Unruh decoherence plays a dominate role in the evolution of the battery with a large acceleration.

\begin{acknowledgments}
We would like to thank Professor Bill Unruh and Professor Philip C. E. Stamp for the discussions on the related work. $\mathrm{SCOAP}^{3}$ supports the goals of the International Year of Basic Sciences for Sustainable Development.
\end{acknowledgments}

\appendix

\section{The detailed derivation of Eq. (4)}
Building upon the results from \cite{Benatti2004}, we will demonstrate a detailed derivation of Eq. (4) which has a GKSL form under the weak-coupling limit. We consider a general interaction between the battery and external coherent field. We write the Hamiltonian of the driven battery in the form of
\begin{equation}
	\label{eq:(A1)}
H_b = \frac{\Omega }{2}(\mathbf{n}\cdot \bm{\sigma}).
\end{equation}
In this work, we chose $ \mathbf{n} = (\cos \varphi ,\sin \varphi ,0)$ and $\varphi $ represents the initial phase of the external driving field.

The definite interaction between the driven battery and environment is given by
\begin{equation}
	\label{eq:(A2)}
	H_I = (\sigma ^{+} +\sigma^{-}) \otimes B=\sigma_{1} \otimes B,
\end{equation}
where the variable $B$ is the quantum scalar field operator $B=\Phi(x)=\sum_{k}[u_k(x)a_k+u_{k}^{\star}(x)a_k^{\dag}]$ of the environmental Hamiltonian. The conditions of  $\sum_k u_k(x)u_{k'}^{\star}(x)\propto \delta_{kk'}$ and $[a_k,a_{k'}^{\dag}]=\delta_{kk'}$ are satisfied. 

The time evolution of the density matrix \(\rho_{tot}\), which describes the state of the entire system, is driven by the total Hamiltonian
\begin{equation}
	\label{eq:(A3)}
	H = {H_S} + {H_I},\	H_S = {H_b} \otimes I_f + I_b \otimes {H_f}
\end{equation}
through standard unitary evolution. Here $I_{b(f)}$ denotes the identity operator of the battery(environment). The initial state is assumed to be factorized: $\rho_{tot}(0)=\rho(0) \otimes \rho_{f} $, where $ \rho(0) $ is the density matrix characterizing the subsystem, and $\rho_{f} $ is the corresponding density matrix for the environment. The environment is assumed to be stationary, meaning $ [H_{f}, \rho_{f}]=0$ (for moving systems, $\rho_{f} \equiv|0\rangle \langle0|$, with $|0\rangle$ being the Minkowski vacuum state).

Accordingly, the dynamics describing the evolution of the battery alone is obtained by tracing over the environmental degrees of freedom. As rigorously proven in \cite{Davies1974,Davies1976markovian}, under the weak-coupling limit between the battery and the environment,the reduced density matrix $\rho(t)=Tr_{f}[\rho_{tot}(t)]$ in the interaction picture satisfies the following evolution equation,
\begin{equation}
	\label{eq:(A4)}
    \frac{\partial \rho(t)}{\partial t}=-\int_0^\infty  {ds} Tr_f\{[H_I(t),[H_I(t-s),\rho(t)  \otimes |0\rangle \langle 0|]]\}.
\end{equation}
The Born-Markov approximation is considered here \cite{Petruccione2002}. The interaction Hamiltonian is $H_I(t)=e^{iH_s t}H_Ie^{-iH_s t}=\sigma_1(t)\otimes B(t)$ where $\sigma_1(t)=e^{iH_b t}\sigma_1 e^{-iH_b t}$ and $B(t)=e^{iH_f t}Be^{-iH_f t}$ are the time dependent operators in the interaction picture.

To express the interaction Hamiltonian, we can obtain $e^{iH_b t}=e^{i\frac {\Omega}2  t}P_{+}+e^{-i\frac {\Omega}2 t}P_{-}$ where $P_{\pm}=\frac{1}{2}(I\pm\bm{n} \cdot\bm{\sigma})$ are the two projection operators. Therefore, the operator $\sigma_1(t)$ is written in the form of
\begin{equation}
	\label{eq:(A5)}
    \sigma_{1}(t)=\sum_{\xi=0, \pm} e^{i \xi \Omega t} \sigma_{1}^{(\xi)},
\end{equation}

where \begin{align}
	\label{eq:(A6)}
	\sigma_{1}^{(0)}=\sum_{k=\pm}P_{k} \sigma_{1} P_{k}=\cos \varphi (\cos \varphi \sigma_1+ \sin \varphi \sigma_2), & \nonumber\\
    \sigma_{1}^{( \pm)}=P_{ \pm} \sigma_{1} P_{\mp}=\frac 12 \sin \varphi(\sin \varphi \sigma_1-\cos \varphi \sigma_2 \mp i \sigma_3).
\end{align}

The Eq. (A4) can be expanded as
\begin{align}
   	\label{eq:(A7)}
   	\frac{\partial \rho(t)}{\partial t} = &\int_0^\infty  ds[\sigma_1(t)\rho(t)\sigma_1(t-s)\otimes\langle 0 |B(t)B(t-s)|0\rangle \nonumber\\
     & -\sigma_1(t)\sigma_1(t-s)\rho(t)\otimes\langle 0 |B(t)B(t-s)|0\rangle + h.c.]
\end{align}
where the term $\langle 0 |B(t)B(t-s)|0\rangle=Tr_f[B(t)B(t-s) |0\rangle \langle 0|]$ is the correlation function of the environment. According to \cite{Petruccione2002}, the reservoir correlation function is homogeneous in time which yields $\langle 0 |B(t)B(t-s)|0\rangle=\langle 0| B(s) B |0\rangle $. To obtain the GKSL form of quantum master equation, we consider the rotating wave approximation which neglect the rapidly oscillating terms. Therefore, the part $\sigma_1(t)\rho(t)\sigma_1(t-s)=\sum_{\xi, \xi^{\prime}=0, \pm}e^{i(\xi+\xi^{\prime})\Omega t} e^{-i\xi^{\prime}\Omega s} \sigma_1^{(\xi)}\rho(t)\sigma_1^{(\xi^{\prime})}$ in Eq. (A7) can be approximately expressed as $\sum_{\xi=0, \pm}e^{i\xi \Omega s} \sigma_1^{(\xi)}\rho(t)\sigma_1^{(-\xi)}$ where the rotating wave approximation of $\xi^{\prime}+\xi=0$ is applied.

This allows us to explicitly evaluate Eq. (A7) and express the result in terms of the Fourier and Hilbert transforms of the environmental correlations, $ \alpha^{(\xi)}(\Omega)=\int_{-\infty}^{\infty} ds e^{i \xi \Omega s}\langle 0|B(s) B| 0\rangle$ and $\beta^{(\xi)}(\Omega)=\int_{0}^{\infty} ds e^{i \xi \Omega s}\langle 0|B(s) B| 0\rangle-\int_{0}^{\infty} ds e^{-i \xi \Omega t}\langle 0|B B(s)| 0\rangle$.

Explicitly, we find,
\begin{align}
  	\label{eq:(A8)}
      \frac{\partial \rho(t)}{\partial t}= & \frac{1}{2} \sum_{\xi=0, \pm} \left\{\alpha^{(\xi)}\left[2 \sigma_{1}^{(-\xi)} \rho \sigma_{1}^{(\xi)}-\sigma_{1}^{(\xi)} \sigma_{1}^{(-\xi)} \rho-\rho \sigma_{1}^{(\xi)} \sigma_{1}^{(-\xi)}\right]\right.  \left.+\beta^{(\xi)}\left[\rho, \sigma_{1}^{(\xi)} \sigma_{1}^{(-\xi)}\right]\right\}.
\end{align}
By using Eqs. (A6)-(A8), we can obtain the GKSL of the quantum master equation like Eq. (4). The response function $\mathcal{G}(\xi \Omega)=\alpha^{(\xi)}(\Omega)$ and the Hilbert transform of the Wightman function $\mathcal{K}(\xi\Omega)=\beta^{(\xi)}(\Omega)$ are also obtained. In this work, the matrix $\bm{a}=(a_{ij})$ can be written as
\begin{equation}
	\label{eq:(A9)}
   \bm{a}= \begin{pmatrix}
		A+C\cos^2 \varphi &C\cos \varphi \sin \varphi &-B\sin \varphi\\
		C\cos \varphi \sin \varphi &A+C\sin^2 \varphi & B\cos \varphi\\
		B\sin \varphi &-B\cos \varphi &A
        \end{pmatrix}.
\end{equation}

	\newpage
	
	{\large \bf Figure Captions}
	
	\vskip 0.5cm
	
	{\bf Figure 1.}
	The asymptotic behaviour of the coherent ergotropy $\xi_{r}^{(s)}$ for the steady state of the UDW battery is monotonically decreased with the scaled acceleration, which provides the evidence that the coherent ergotropy can be viewed as an effective witness to Unruh thermalilty.

	\vskip 0.5cm
	
	{\bf Figure 2}
	
The dynamics of the coherent ergotropy $\xi_r$ and incoherent ergotropy $\xi_i$ is illustrated as function of the external field phase and scaled proper time when the acceleration is chosen to be a certain value $a=0.2$. (a) The time-dependent evolution of the coherent ergotropy is shown. (b) The effect of the field phase on the incoherent (red dashed line) and coherent ergotropy (black solid line) is demonstrated at $\tau=0.7$.
				
	\vskip 0.5cm
	
	{\bf Figure 3.}
	
	The behavior of the coherent ergotropy ${\xi _r} $ is plotted as function of the scaled acceleration and driving field phase at a certain proper time $\tau  = 0.3 $

\newpage
	\begin{figure}
		\centering
		\includegraphics[width=1\textwidth]{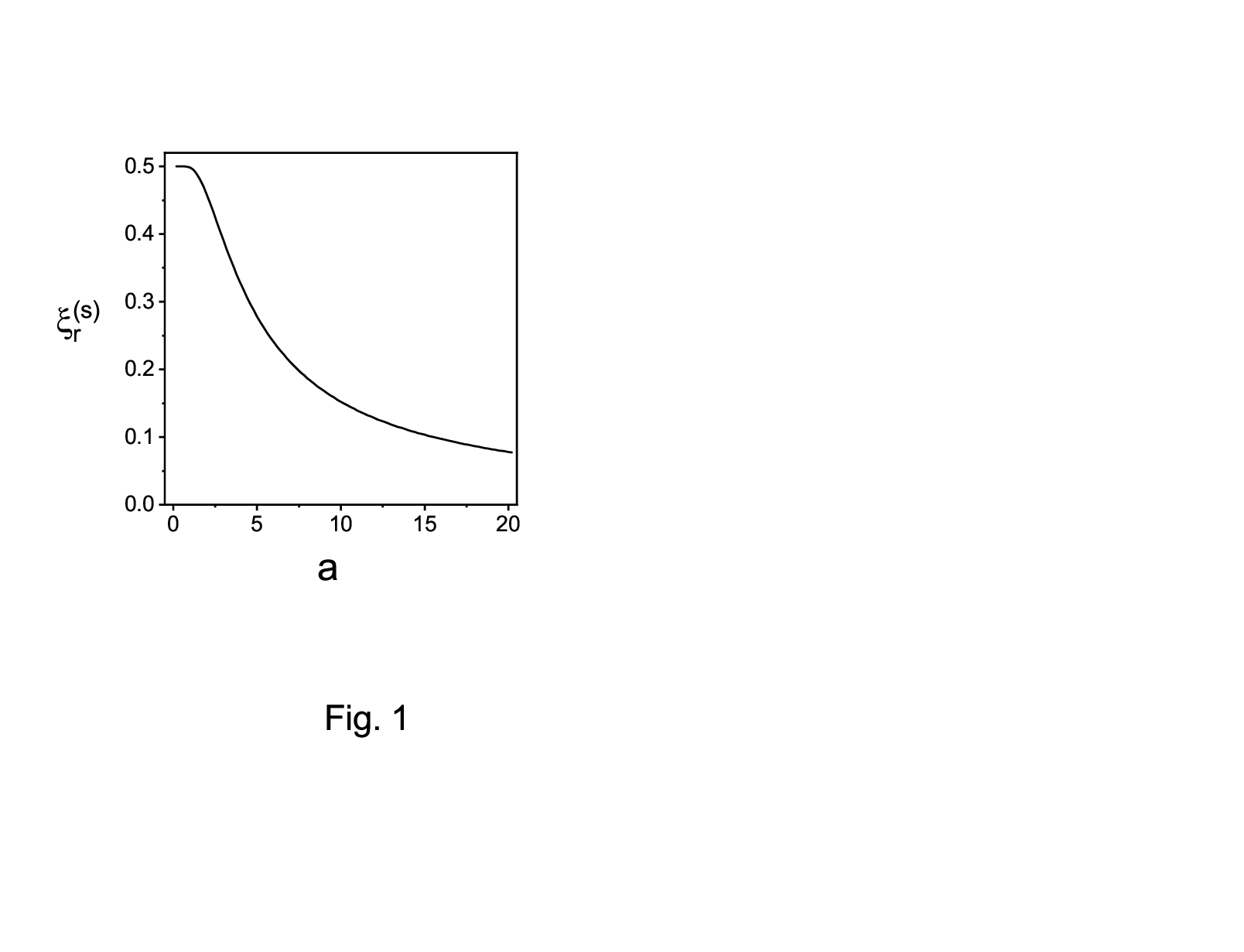}
	\end{figure}
	
	\begin{figure}
		\centering
		\includegraphics[width=1\textwidth]{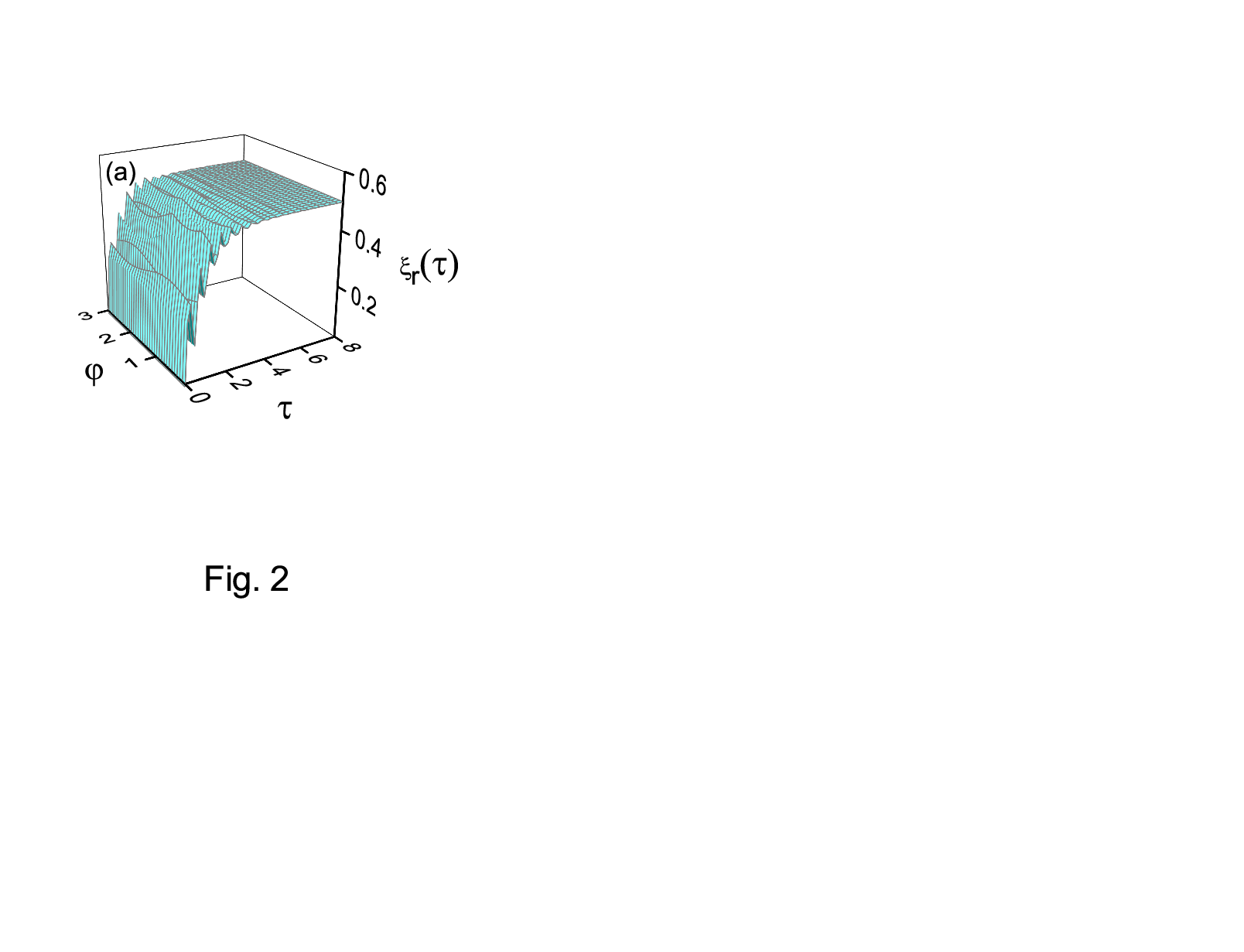}
	\end{figure}
	
	\begin{figure}
		\centering
		\includegraphics[width=1\textwidth]{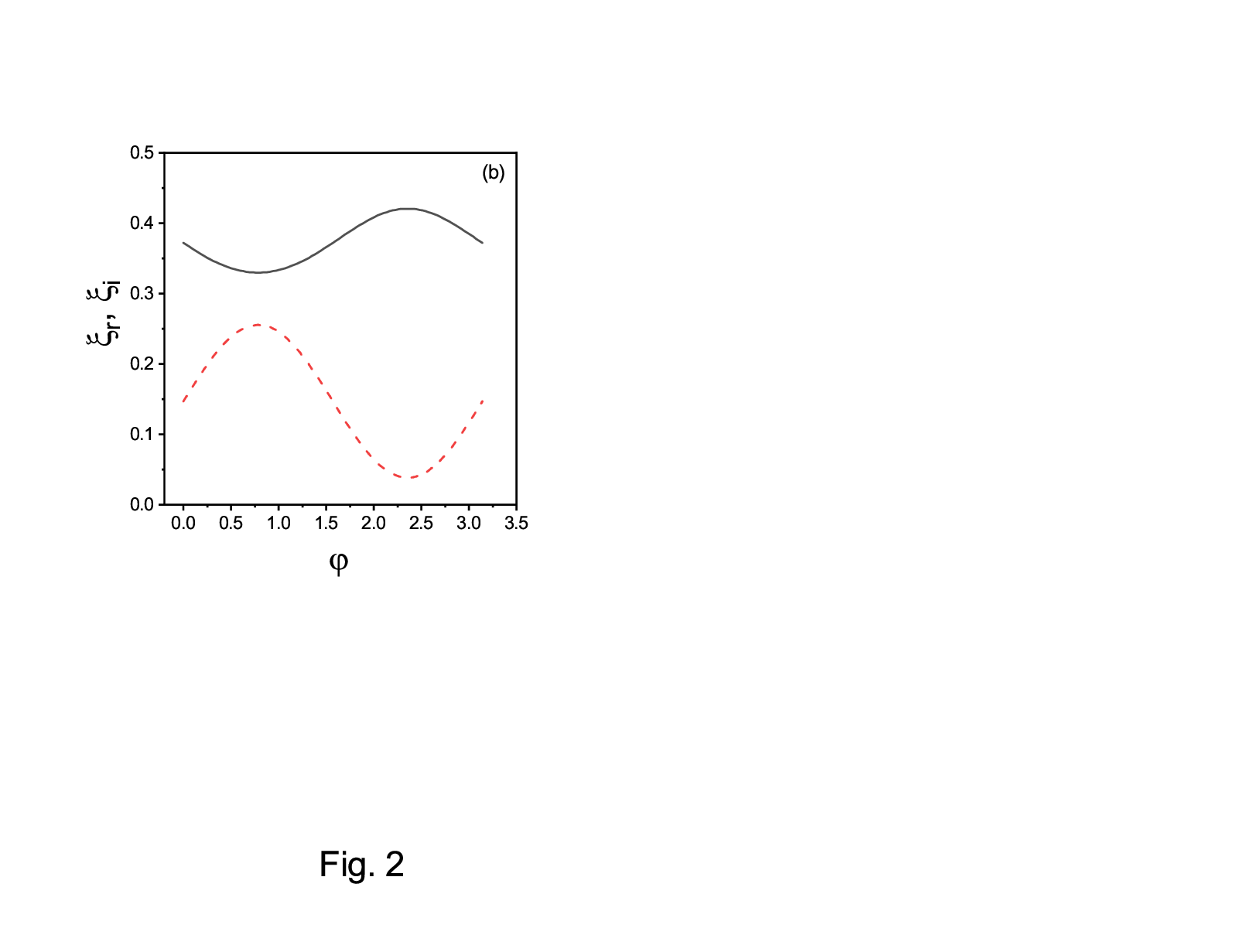}
	\end{figure}
	
	\begin{figure}
		\centering
		\includegraphics[width=1\textwidth]{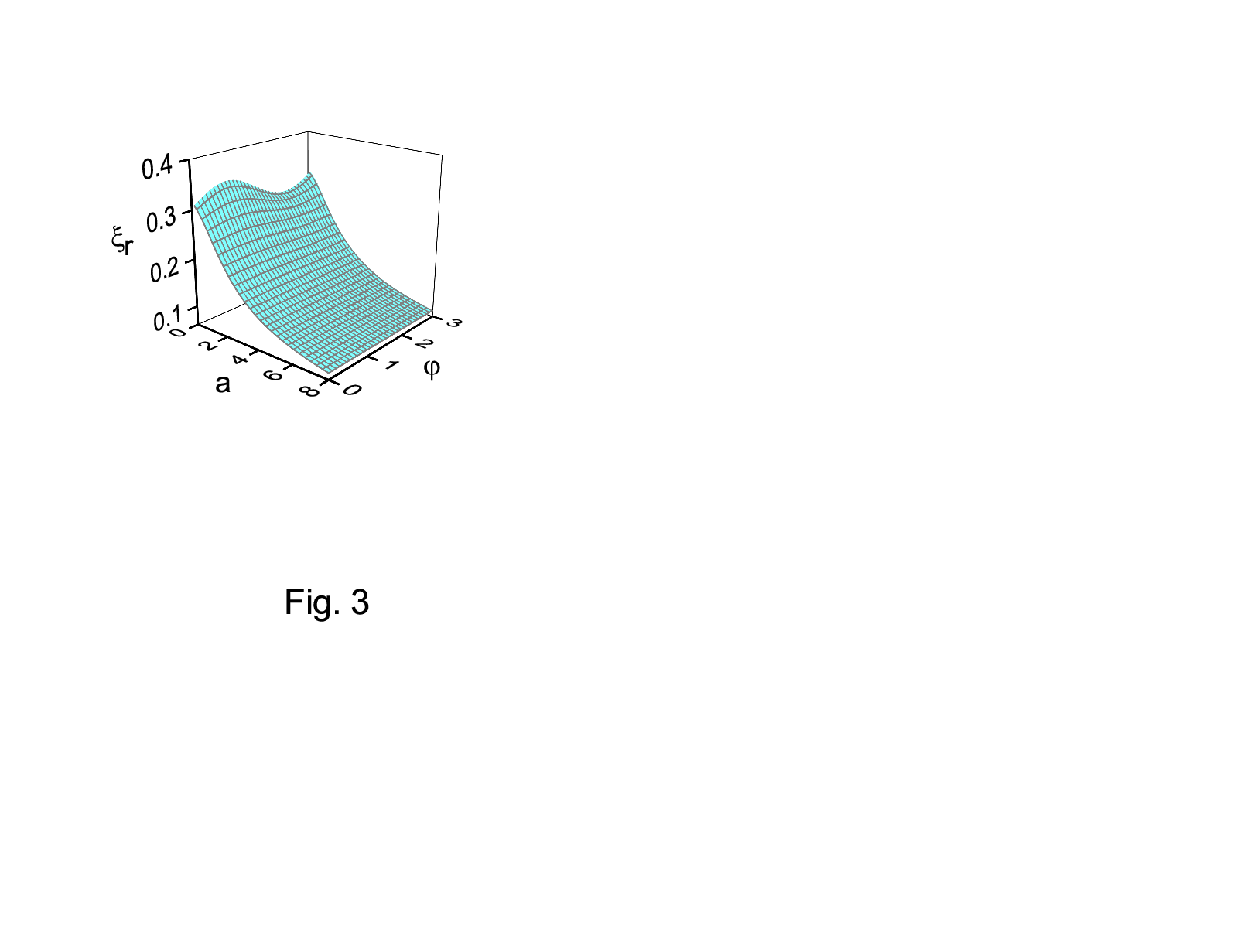}
	\end{figure}

\end{document}